# Superconducting Nanowire Fabrication on Niobium Nitride using Helium Ion Irradiation

Glenn D. Martinez[1,2], Drew Buckley[1,2], Ilya Charaev[1], Andrew Dane[1], Douglas E. Dow[2], Karl K. Berggren[1]

[1]Research Laboratory of Electronics, Massachusetts Institute of Technology (MIT), Cambridge, MA, USA

[2]Dept. of Electrical and Computer Engineering, Wentworth Institute of Technology, Boston, MA, USA

*Abstract*— Superconducting devices are prone to reduced performance caused by impurities and defects along the edges of their wires, which can lead to local current crowding. In this study, we explored the use of helium ion irradiation to modify the lattice structure of the superconducting material to change its intrinsic properties. The process will allow us to directly pattern devices and potentially improve the quality of the nanowires. To achieve this, we used the ion beam from a scanning helium ion microscope (HIM) to localize damage on a superconducting material to create a nanowire. Two experiments were performed in this study. First, a range of helium ion doses was exposed on a niobium nitride (NbN) microwire to determine the estimated dose density to suppress superconductivity. Using the results of this first experiment, nanowires were patterned onto a microwire, and the current-voltage characteristics were measured for each sample. Our results showed that helium ion irradiation is an effective resistless fabrication method for superconducting nanowires.

## I. INTRODUCTION

Superconducting nanowire single photon detectors (SNSPDs) have played a prominent role in the field of quantum information science. SNSPDs are devices that count single photons within the infrared wavelength range. Applications of these devices include space-to-ground communications and optical quantum computing [1]. The fabrication of SNSPDs involves processes including film deposition, lithography, and etching. One issue is maintaining the quality of the superconducting thin films (<10 nm). The quality of the material directly impacts device performance [2]. During the lithography process, a sample is coated with a chemical resist to create nanoscale features. Removing this resist requires the coated sample to be submerged in a developer chemical. A study was conducted to determine the effects of the development process on a common material used for SNSPDs, niobium nitride (NbN) [3]. It was shown that the developer can decrease the thickness of NbN and can form clusters of niobium. These clusters impact subsequent processes, such as etching, and device performance.

To improve the sensitivity of the SNSPD, narrow nanowires are needed [4]. However, the fabrication of these nanowires at the smaller scale starts to introduce imperfection along the edges of the nanowires. These defects can suppress the critical current $I_c$, or the maximum current before the superconducting material switches to the normal state [5]. The suppression occurs due to an increased current density on the rough edges of the nanowire. Also known as current crowding, this effect negatively impacts performance for SNSPDs [6]. Reducing the number of steps in the fabrication process can potentially lessen the probability of damaging the material.

Recently, a new tool called the helium ion microscope (HIM) was developed, which uses helium ions for imaging [7]. In addition to imaging, some applications of this tool include helium ion lithography and milling [8,9]. Another effect of interest in this study is directly writing the device without the use of resists. Applying enough damage with He$^+$ ions can change the property of the superconductor to be resistive or even insulating at temperatures when the unexposed material would be superconducting [10]. Following the same principle, the He$^+$ ion beam was used to fabricate narrow, insulating barriers to develop Josephson junctions on YBa$_2$Cu$_3$O$_7$ (YBCO) and MgB$_2$ [11,12]. This method was able to limit the number of processes, such as lithography, needed to fabricate as well as improved device performance [13].

In this paper, we present a resistless fabrication method for creating superconducting nanowires on a material used in SNSPD fabrication, NbN, by using the HIM's ability to localize damage. An example device is shown in Fig. 1. By using the He$^+$ ion beam to damage the superconducting material, dose tests were performed to understand the effects of He$^+$ ion irradiation on NbN. Based on these results, we optimize the HIM's parameters to develop high-resolution nanowires and analyzed their current-voltage characteristics.

## II. METHODS

The Zeiss Orion Plus helium ion microscope was equipped with a built-in 16-bit raster scan pattern generator and an external Raith Elphy Multibeam. The former scans horizontally with the beam turning on when crossing an area that needs to be exposed and turns off otherwise. The latter is a pattern generator that is capable of vector scanning. Vector scanning only scans directly on the regions that need to be exposed, which facilitates the patterning of complex layouts. Silicon nitride (SiN) was used as the substrate with a 5 nm thin film of NbN sputtered using an AJA Sputtering System. The procedure for deposition was described in detail in [14]. The thin film had a critical temperature 9.7 K. Contact pads connected by a 2 µm wire were fabricated with electron-beam lithography. The samples were coated in positive electron-beam resist ZEP 520A and developed in O-xylene. Reactive ion etching in CF$_4$ was used to transfer the pattern. Parameters were detailed in [15].

*A. Dose Test*

To determine the estimated dose to modify the NbN thin film and suppress superconductivity, a He$^+$ ion beam was drawn across the prefabricated microwire using the built-in pattern generator. An area exposure was used with the dimensions being 3µm × 50nm. Doses for each exposure ranged from $1\times10^{15}$ to $1\times10^{20}$ ions/cm$^2$. For this experiment, the HIM was set to have an acceleration voltage of 30 kV. To account for patterning with large doses and to reduce write time, the He+ ion beam current was increased. This can be done by increasing the helium pressure and using a larger aperture. For this experiment, a 20 µm aperture was selected as it provided the most current from the beam and the helium pressure was set to $8\times10^{-6}$ Torr to achieve a beam current of 6-pA. Also, the working distance was set to 6.8 mm, as we observed this setting to result in the highest resolution patterns.

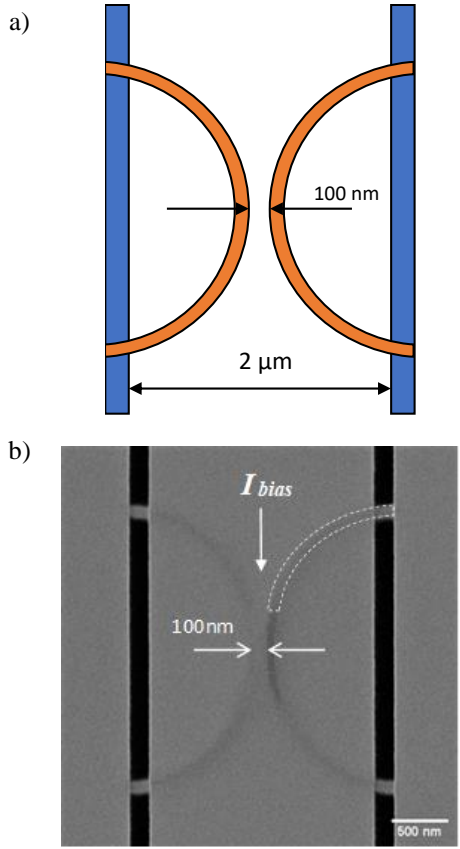

Fig. 1. a) Layout of the 100 nm wide nanowire to be patterned with He+ ion irradiation on a 2 µm wide NbN wire. Blue rectangles represent area patterned by electron-beam lithography while the orange arcs represent area exposed with He+ ions. b) HIM image of a 100 nm nanowire patterned with the He+ ion beam. A dotted outline was used to indicate one part of the He-exposed region. I bias represents direction of current when testing the device. The region defining the edges of the nanowire was exposed with a dose of $1\times10^{17}$ ions/cm$^2$ and showed no visible swelling.

### B. Nanowire Fabrication

The Raith Elphy Multibeam pattern generator was needed for the fabrication of the nanowires. In Fig. 1(a), the nanowire layout was a curved line with a center width of 100 nm. This would be patterned onto the existing microwire structure that was fabricated using electron-beam lithography. The HIM's acceleration voltage, aperture, and working distance remained unchanged. As shown in Fig. 1(b), a nanowire was successfully patterned.

### C. Testing

Once the exposures were completed, the chips were placed onto a dedicated PCB. The contact pads of each sample were wire-bonded to the output channels of the PCB. The samples were cooled to 4.2 K inside a liquid helium Dewar with current-voltage characteristics and resistance being measured. A low-noise voltage source was connected through a 100 kΩ resistor to produce a current to the device via 2-point probe. The signal from the device then passed into a bias-tee to filter the noise. Software controlled the voltage source as well as recorded the data points from the voltmeter.

## III. RESULTS AND DISCUSSION

### A. Dose Test Results

To determine if superconductivity was suppressed by He+ ions, the IV-curve and resistance of the microwire were measured. In Fig. 2(a), critical current $I_c$ is plotted as a function of ion dose. The critical current decreases with increasing dose. For the doses from $5\times10^{17}$ to $1\times10^{20}$ ions/cm$^2$, $I_c$ was not observed for tested structures. Data for measured resistance is shown in Fig. 2(b). In contrast to the critical current, the resistance was proportional to the dose. Although microwires have zero resistance below their critical temperature, small resistance was observed for samples exposed at lower doses possibly from the He$^+$ ion irradiation. At $5\times10^{17}$ ions/cm$^2$ and above, the resistance started to drastically increase with a maximum resistance of $1\times10^7$ Ω/sq.

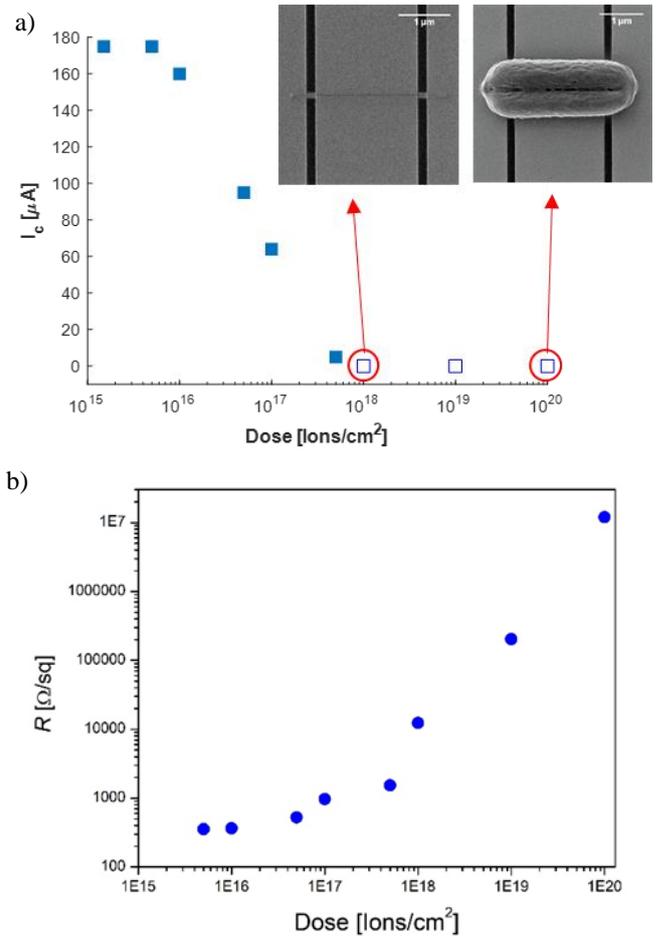

Fig. 2. Critical current $I_c$ and resistance R as functions of He+ ion exposure. The sample was exposed with the following HIM parameters: 30 kV acceleration voltage and a beam current of 6 pA. (a) Critical current of the microwires exposed with He+ ion doses ranging from $10^{15}$ to $10^{20}$ ions/cm$^2$. The hollowed square represents no critical current measured at that dose. HIM images of a damaged 2 µm microwire exposed with $1\times10^{18}$ and $1\times10^{20}$ ions/cm$^2$ doses are displayed to showcase the physical damage on the material. b) Resistance as a parameter to determine the effect of the dose. As shown in the plot, resistance increased with the dose, indicating that He+ ions suppressed superconductivity at 4.2 K.

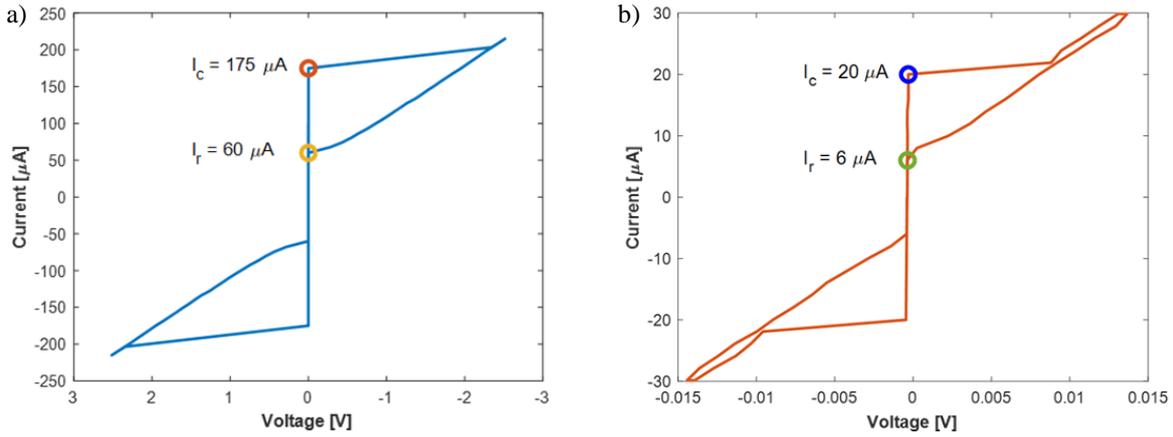

Fig. 3. Current-voltage characteristics of an unexposed 2 µm wire (a) and a patterned 100 nm width nanowire (b). The critical current $I_c$ and retrapping current $I_r$ measurements are marked within the circle with their respective value. Notice that the $I_c$ and $I_r$ in plot (b) have a significant reduction in current for the patterned nanowire wire compared to the measurements in plot (a). Since $I_r$ and $I_c$ are directly proportional to the width of the nanowire, a nanowire was successfully patterned. Comparing the ratio of these two values with the unexposed nanowire and patterned nanowire samples showed minor differences between the two.

A reason why this occurred can be seen through HIM images of the dose tests. In Fig. 2(a), the damage was visible with a dose of $1\times10^{18}$ ions/cm$^2$ and became worse at $1\times10^{20}$ ions/cm$^2$ since there was visibly swelling around the intended exposed area. The result can be explained by a study that was conducted to observe the effects of He$^+$ ion irradiation on silicon and copper substrate at varying doses [16]. A high dose applied to the sample increases He$^+$ ion accumulation and produces sub-surface micro and nano-bubbles. Continuing to dose on an overexposed area will lead the swelling to pop and destroy the material.

### B. Nanowire Results

The data from the dose test was used to roughly inform the nanowire fabrication process. Experimentally, a dose of $1\times10^{17}$ ions/cm$^2$ was shown to have no visible swelling and a resulting pattern true to the layout's dimensions. The dose was lower than expected when comparing to the initial dose test results. One possible explanation can be due to the larger exposure area of the nanowire pattern. He$^+$ ion collisions possibly overlapped, which will increase the likelihood of heavier damage at lower doses.

The current-voltage measurements have been done to determine the critical and retrapping currents of micro and nanowires. Fig. 3(a) and Fig. 3(b) are the current-voltage measurements for an unexposed 2-µm wide microwire and a patterned 100-nm wide nanowire. The critical current, $I_c$ of all structures was associated with the well-pronounced jump in the voltage from zero to a finite value corresponding to the normal state. Retrapping current $I_r$ (4.2 K) is the current at which the wires return from the resistive state back to the superconducting state when the current decreases from $I > I_c$ to zero. Comparing these two current values for the unexposed microwire with the patterned nanowire, we observed a decrease in $I_c$ from 175 µA to 20 µA and a decrease in $I_r$ from 60 µA to 6 µA. This was an indication of the patterned nanowire reducing the superconducting area of the microwire. In addition to the reduction of $I_c$ and $I_r$, the transition from the superconducting state to the normal state is smooth. Otherwise, the transition profile would have instances of voltage being measured before entering the normal state due to part of the region on the microwire being irradiated with He$^+$ ions.

The reproducibility of the results was checked by patterning additional samples with the previous experiment's conditions. The sputtering of NbN on SiN and the electron-beam lithography process followed the same procedure referenced in the Methods section. The same 100 nm nanowire pattern and a dose of $1\times10^{17}$ ions/cm$^2$ were used in this experiment. Three exposed samples were measured, and each one was observed to have a decrease in both $I_c$ and $I_r$.

## IV. CONCLUSION AND FUTURE DIRECTIONS

In conclusion, we have successfully fabricated nanowires by directly writing the pattern onto the NbN thin film. By damaging the microwires at various doses, we optimized the HIM's parameters to suppress superconductivity and to produce high-resolution nanowire patterns. By comparing an unexposed microwire and a patterned nanowire, it was confirmed that the method was capable of device fabrication. This knowledge can be expanded to the fabrication of devices that use superconducting nanowires, specifically superconducting nanowire single-photon detectors (SNSPDs).

## ACKNOWLEDGMENTS

The authors would like to thank Jim Daley and Mark Mondol of the MIT Scanning-Electron Beam Lithography Facility for valuable discussion regarding the operation of the helium ion microscope and fabrication. Lastly, the authors acknowledge Skoltech for supporting this project.